%% file: pwd.tex
\documentclass{llncs}

\usepackage{amsmath}
\usepackage{amssymb}
\usepackage{xspace} 
\usepackage{mathptmx}
\usepackage{stmaryrd}
\usepackage{mathtools}
\usepackage{tikz}
\usepackage{listings}

\input{linkslang}

\newcommand{\li}[1]{\lstinline{#1}}

\title{Curating Covid-19 data in Links}
\titlerunning{Curation in Links}

\author{
Vashti Galpin\inst{1}\orcidID{0000-0001-8914-1122}\and
James Cheney\inst{1,2}\orcidID{0000-0002-1307-9286}}

\institute{University of Edinburgh, UK\\
\email{\{Vashti.Galpin,James.Cheney\}@ed.ac.uk}
\and
The Alan Turing Institute, UK}

\begin{document}
\maketitle

\begin{abstract}
Curated scientific databases play an important role in the scientific
endeavour and support is needed for the significant effort that goes
into their creation and maintenance. This demonstration and case
study illustrate how curation support has been developed in the
Links cross-tier programming language, a functional, strongly typed
language with language-integrated query and support for temporal
databases. The chosen case study uses weekly released Covid-19
fatality figures from the Scottish government which exhibit updates
to previously released data. This data allows the capture and query
of update provenance in our prototype. This demonstration will
highlight the potential for language-integrated support for
curation to simplify and streamline prototyping of web-applications in
support of scientific databases.
\end{abstract}

\section{Introduction}

Curated scientific databases take significant human effort to develop and
then to maintain \cite{BuneBCTV:08}. These databases are crucial in providing
knowledge bases for science to proceed and users of the data need to be
able to trust in their contents. One way to support this trust is for
users to be able to review provenance information.

We consider update provenance \cite{BuneBCTV:08} as a specific use
case, based on the fact that regularly released datasets can include
updates to previously released datasets, and our case study is based
on data that has been released weekly by the Scottish government
on Covid-19 deaths \cite{NRS:21}.

We are developing a prototype curation interface using
Links\footnote{\url{https://links-lang.org/}}, a
cross-tier programming language with language-integrated query that
is being extended with experimental temporal database features
\cite{CoopCLWY:06}.

Both the prototype and Links's temporal features are works-in-progress,
and as we develop the interface, we are considering the more general
issues that would apply to any curated scientific databases where
update provenance is to be captured using temporal databases.

\section{Links}

Links is a cross-tier programming language developed at the University
of Edinburgh \cite{CoopCLWY:06}.  It is \emph{cross-tier} in the sense that
it allows a developer to write a single type-checked program which
that can then be executed efficiently on multi-tier Web architectures.
Typically, a web application consists of three distinct applications
coded in different languages: HTML and JavaScript for the browser, a
different language for the application logic and SQL for the database
queries, leading to complex coordinating issues. Links provides a single
language in which to implement these three aspects, thus simplifying the
development process. For our curation interface, the built-in
translation to JavaScript completely abstracts away the details and
there is no JavaScript code in the source Links program.

Furthermore, Links supports \emph{language-integrated query} whereby
database queries can be written in Links itself resulting in safe
and efficient queries. In particular when writing database-related
code, the support provided by type inference for row types in Links
makes referring to the fields of interest in a table while abstracting
from the others straightforward. This leads to compact and readable
programming style.  Links requires the types of the database tables
to be declared explicitly in the program, which enables checking
of language-integrated queries and supports the temporal query
transformations. This adds some work at the beginning and if there
is schema evolution, work is needed re-synchronize these declarations.
Support for schema evolution in Links is a future area of research.
Links currently supports PostgreSQL, MySQL 5.x and
SQLite 3. Links covers a wide range of SQL but some aspects are
ongoing research such as support for set and bag semantics
\cite{RiccRC:21}.

In terms of use for curated scientific databases, IUPHAR/BPS Guide
to PHARMACOLOGY (GtoPdb) \cite{ArmAFDx:20} has been reimplemented
in Links as a case study to demonstrate that that Links is suitable
for this task both in terms of functional correctness and performance
\cite{FowlFHSJ:20}. However, because of its size and complexity,
GtoPdb is not a good candidate for exploring the strengths and
weaknesses of Links's new temporal support; instead in this
demonstration we present the current Covid-19 curation prototype,
which is the next step  in developing curation functionality for
this type of database in Links. We consider both the interface as the
end product, and how the development of the interface is supporting by
the features of Links.

\section{Temporal databases}

\begin{figure}[t]
\hspace*{0.7cm}
\includegraphics[width=11cm]{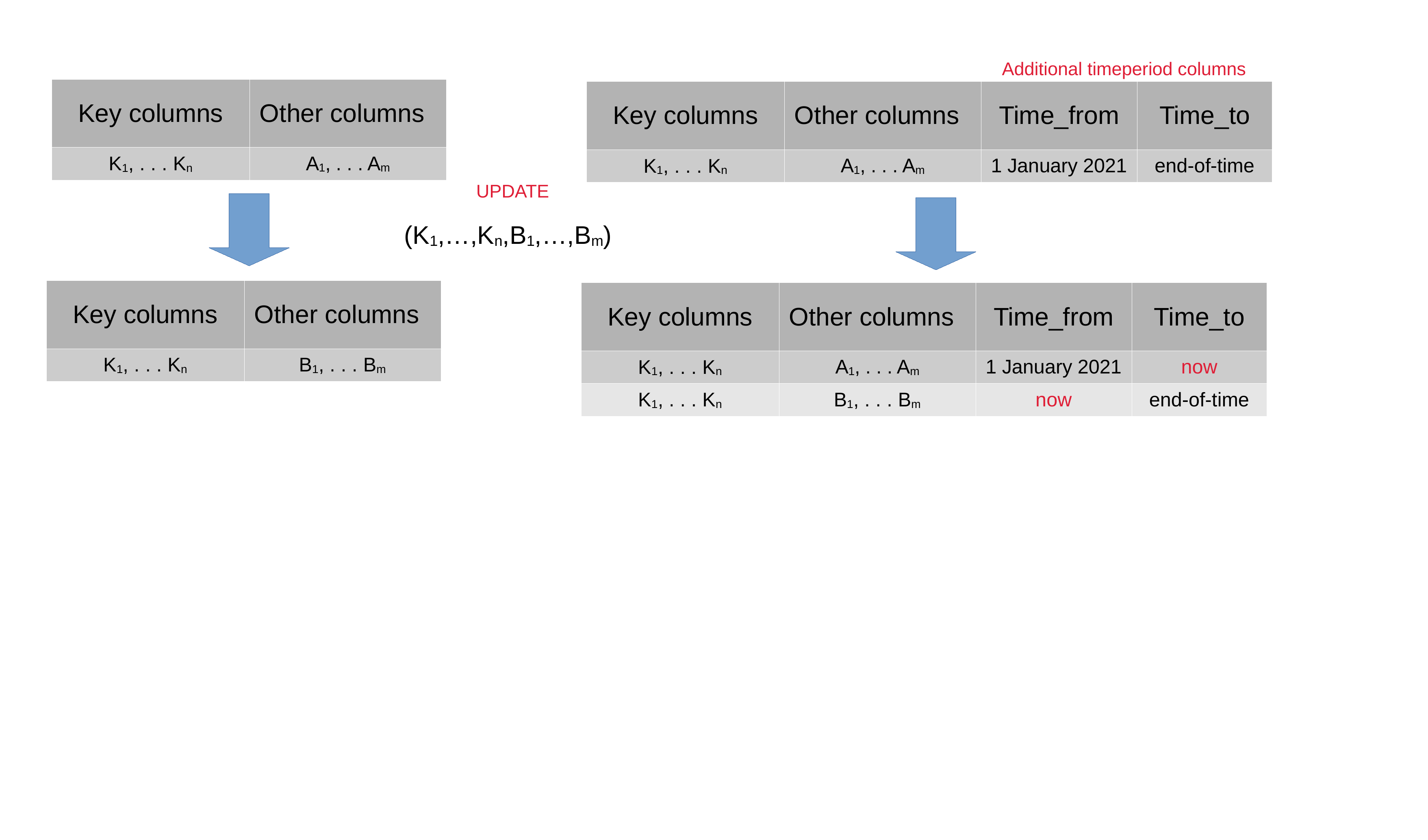}
\vspace*{-3.3cm}
\caption{Standard database update (left) compared with temporal database
update (right)} \label{fig-tdb}
\end{figure}

Temporal databases provide the ability to record when a row in a table
is valid, either with respect to the database itself using a
transaction time period, or with respect to the real world using a valid
time period \cite{JensJS:09}. We can use this functionality to track
the validity of the data items that are updated. Figure~\ref{fig-tdb}
illustrates how an update differs between a standard database on
the left, and a temporal database on the right. In the former, the
data is replaced and the previous values are lost. However, in the
temporal case, additional columns to record the start and end of
the time period of validity form part of the key and allow previous
values to be recorded. The interpretation of the validity is dependent
on the particular application.

Although temporal extensions to SQL have been standardized
\cite{KulkKM:12}, many current popular relational database
implementations have no built-in support for temporality, although
it is possible to implement temporal tables by the use of explicit
additional fields and complex hand-generated SQL queries \cite{SnodS:99}.
In comparison, Links now provides support for transaction-time
tables and valid-time tables, allowing both sequenced (taking time
period into account) and non-sequenced queries over these tables.
This is achieved by interacting with a standard relational database
such as PostgreSQL, using the approach currently provided by Links
for generating standard queries and translating temporal queries
into standard queries as described above \cite{SnodS:99}, thereby
avoided the need for the application developer to generate these
complex SQL statements. An example is given in the next section.

\section{Curation functionality}

To investigate how to develop curation functionality in Links, we
choose a dataset that raised questions of interest around updating of
data. To set the context for discussion of the type of queries that can
be considered, we first provide more information about the data we used
in implementing the prototype.

\subsection{The case study}

\begin{figure}[t]
\vspace*{-0.5cm}
\hspace*{1.5cm}
\includegraphics[width=9cm]{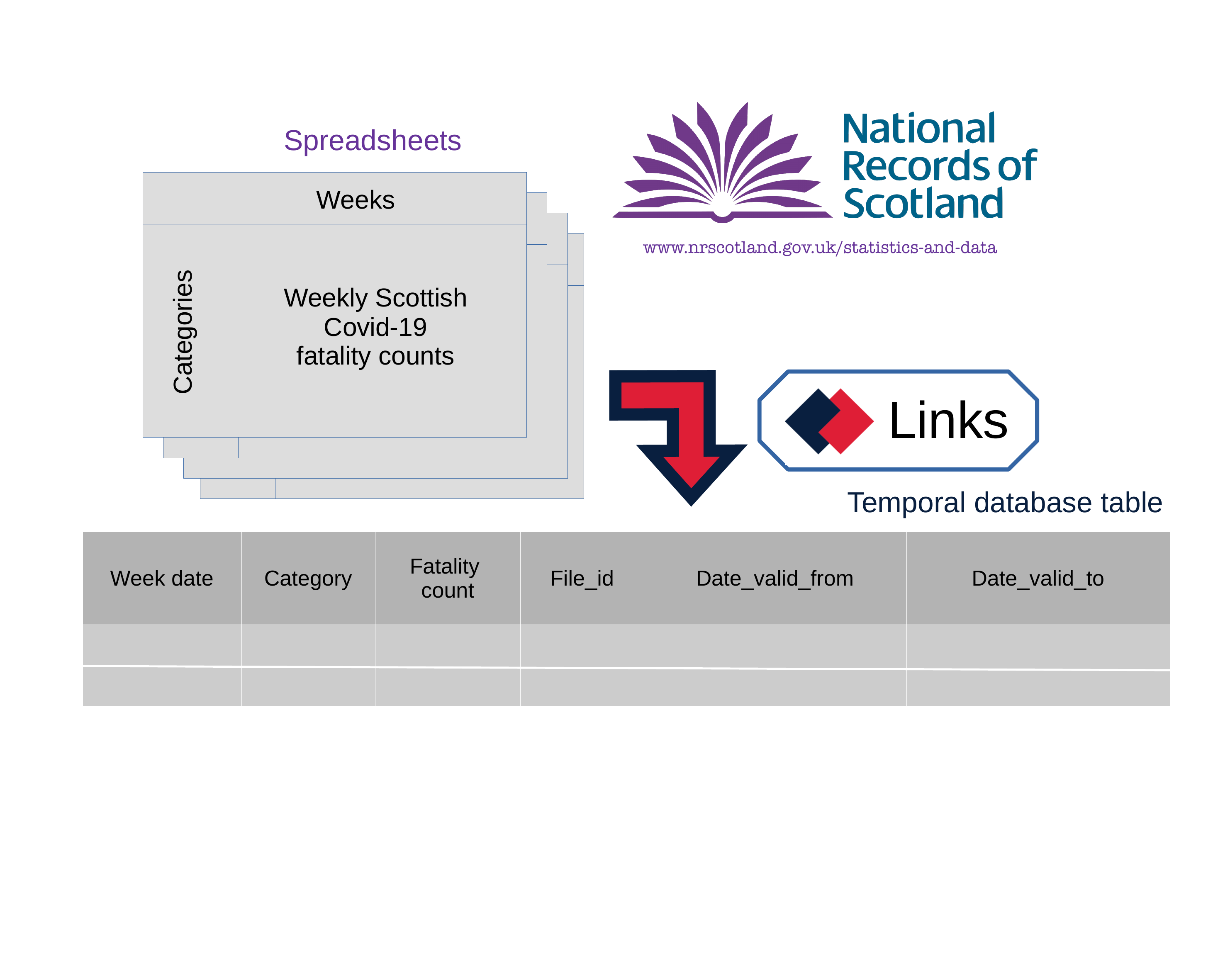}
\vspace*{-2cm}
\caption{Uploading new data} \label{fig-sstotdb}
\end{figure}

We identified data that was released weekly by the Scottish Government
through the National Records of Scotland website \cite{NRS:21}, as suitable and
interesting for the prototype exploration. Each week, since early in the
pandemic a spreadsheet (as a CSV file and Excel document) has been
released (amongst other data) with counts of Covid-19 fatalities for
individual weeks and data categories including sex, age, health board,
local authority, and location. An example of a data category is ``Location of
death'' and its subcategories are ``Care Home'',
``Home/Non-institution'', 
``Hospital'' and ``Other institution''. Our application supports upload of
this CSV data and its transformation into a temporal table as
illustrated in Figure~\ref{fig-sstotdb}.

\subsection{Provenance-based queries}

One of the aims of this research is to identify the queries that
are interesting for update provenance. An obvious query is how an
individual data item has changed over time, but other potential
queries include finding out whether one category of data has changed
more often than another. Dependency is also relevant: are the changes
of a data item associated or correlated with changes in other data.

\pagebreak

Examples of queries that the interface can support (or will 
support in the future):
\begin{enumerate}
\item What was the number of female fatalities in the week of 20 April
2020 when that data was first uploaded 
\item What is the current number of female fatalities in the week of 20 April
2020? 
\item What is the range of female fatalities in the week of 20 April
2020 over all uploads? 
\item Which updates to female fatalities were rejected and when?
\item Have there been more updates to female fatalities or male
fatalities considering all weeks?
\item Which health board has seen most updates in the first six
months of the pandemic?
\item How are updates for the Lothian health board figures correlated
with those for the local authority of Edinburgh?
\end{enumerate}

\subsection{The prototype curation interface}

\begin{figure}[t]
\hspace*{-0.7cm}
\includegraphics[width=7.5cm]{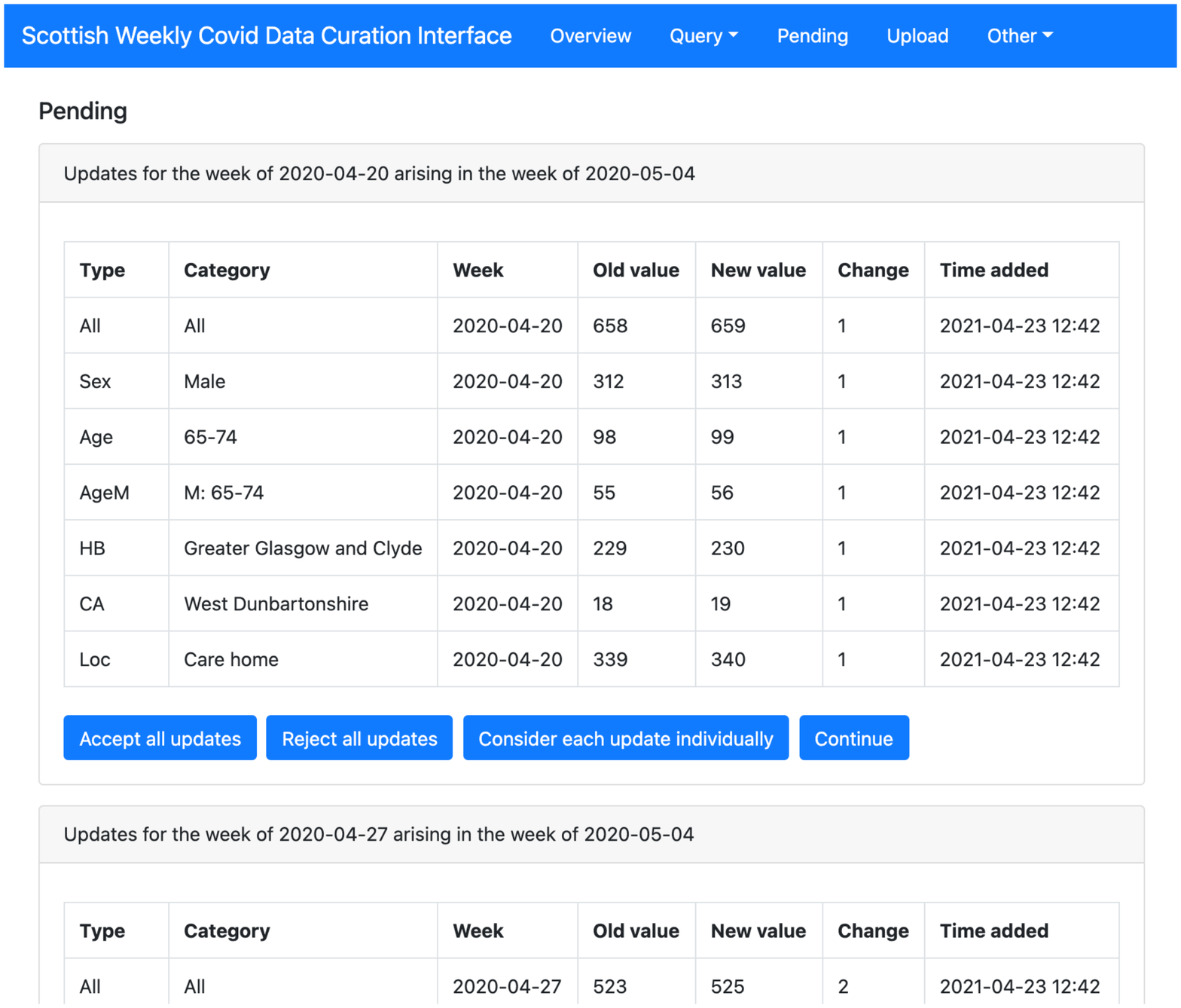}
\hspace*{-0.5cm}
\includegraphics[width=5.73cm]{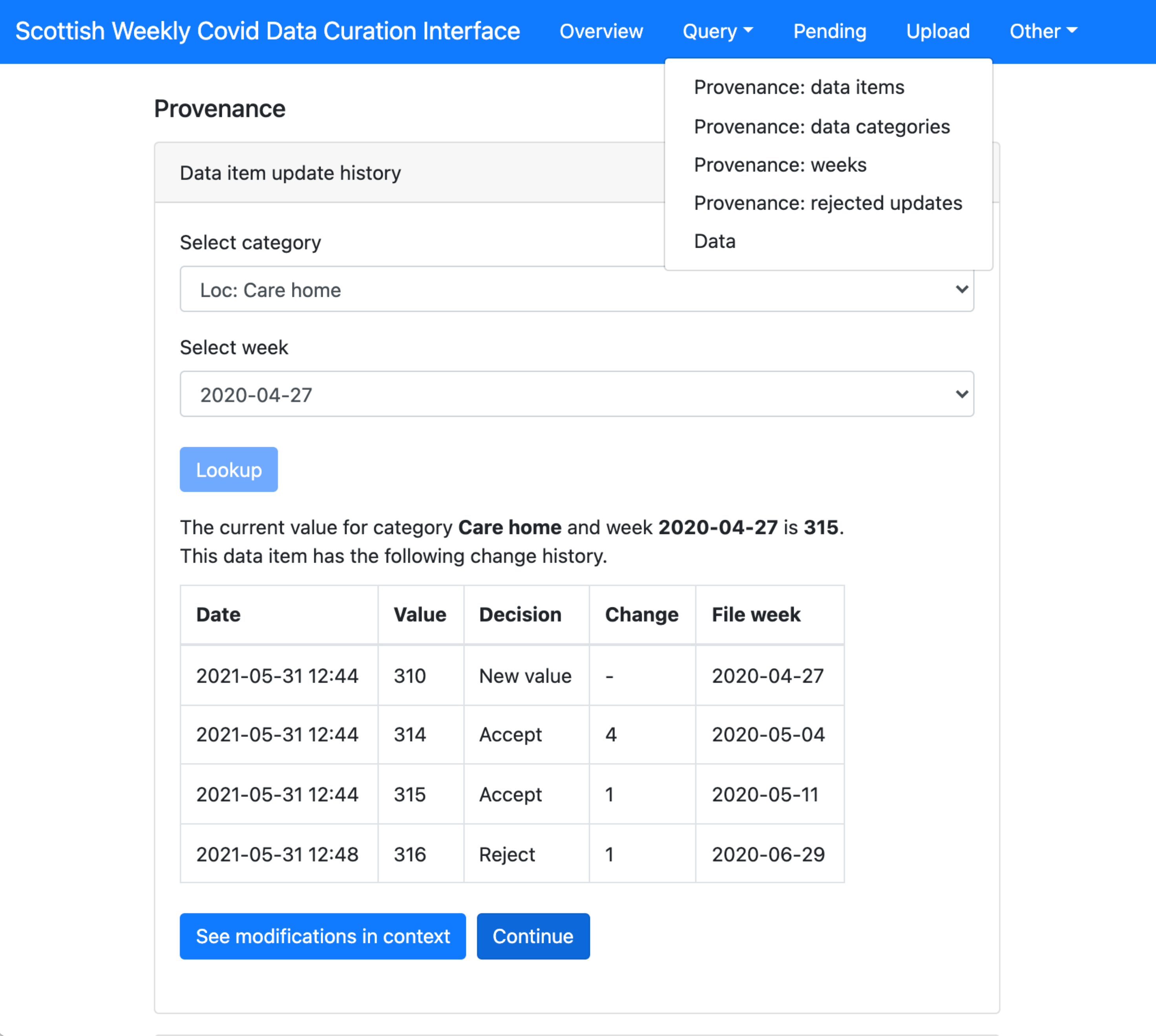}
\caption{Screenshots: decision making (left) and provenance (right)} \label{fig-dec}
\end{figure}

The interface has been developed using the Bootstrap HTML, CSS and JavaScript
library.
We present two screenshots illustrating the functionality of the
prototype interface in Figure~\ref{fig-dec}. The
left-hand figure shows updates for a specific week that have arisen in a
subsequent week. By grouping them together, a user is able to assess the
consistency of this update\footnote{The dataset has the feature that it
contains counts for categories that subsume other categories. We decided
to capture all categories rather than just the
minimal ones to support checking for inconsistencies.}. These
updates can be accepted or rejected together, or each can be considered
individually in the context of other updates to that data item.

The following shows the Links code which updates the
main table of fatalities from a table of accepted updates for counts
relating to the Lothian health board.

\begin{lstlisting}
for (x <-- AcceptedUpdates)
      [update sequenced (y <-v- CovidDeaths)
        between (valid_from_date,forever())
        where (x.week==y.week && x.category=="Lothian")
        set (count = x.new_value, file_id=x.new_id)]);
\end{lstlisting}

The double arrows, \li{<--} and \li{<-v-}, indicate iteration over the
rows of the named tables, with \li{<-v-} indicated that the table is
a temporal table.
The keyword \li{sequenced} indicates that this update should result in
the modification of one record and the insertion of a new record as
illustrated in Figure~\ref{fig-tdb}. To write this query in SQL for a
non-temporal database would be more complex, requiring explicit updates
and insertions.

The right-hand screenshot in Figure~\ref{fig-dec} shows a provenance
query on a single data item.  For the updates on this data item,
it is possible to see the updates in the context of other updates
that occurred in the same week.  Other provenance queries can be
done on data categories (such as queries 5, 6 and 7 above), and a
similar approach can be applied to weeks. The final menu item allows
for queries such as query 4.

The prototype has taken around 120 hours to develop to date,
approximately 1 person month of effort by the first author, who was
not previously familiar with Links. It consists of around 1800 lines
of Links code, as well as supporting CSS and JavaScript code. Reuse of
existing code will make the implementation of additional queries less
time-consuming.


\section{Conclusion}

We have developed a prototype curation interface in Links to demonstrate
and investigate curation functionality for a selected dataset with
interesting features relating to data update. 
Links, as a cross-tier programming language, has provided a typed
and functional language for implementing all of the appearance of
the web interface (using the Bootstrap library), the logic of the
interface and querying of the database.

By using Links, we have avoided the need to code temporal features
in SQL for databases without temporal features, to transition to a
database such as MariaDB \cite{MariaDB:18} which has temporality, or to deal
with the integration errors that occur during conventional web
development using SQL.

Another option would be to use a temporal middleware/stratum
translator that maps temporal SQL queries to plain ones \cite{TorpT:09}.
This requires the same kind of information about the database schema
that Links requires, but does not provide the advantages of language
integration and type checking that we get from Links.

We have successfully created an interface that supports a number
of queries and further work involves determining how to support the
generation of a curation interface with an arbitrary database schema
where selected tables have temporal features.

\paragraph{Acknowledgements}
This work was supported by ERC Consolidator Grant Skye (grant number
682315) and by an ISCF Metrology Fellowship grant provided by the
UK government's Department for Business, Energy and Industrial
Strategy (BEIS). We thank Simon Fowler for his assistance in using the
temporal extensions to Links.


\bibliographystyle{splncs04} \bibliography{pwd}

\end{document}

%% file: linkslang.tex
\lstdefinelanguage{Links} {morekeywords={%
        universe,
        weight},
sensitive=false,
morecomment=[l]{//},
morecomment=[s]{/*}{*/},
morestring=[b]",
mathescape=true,}

\usepackage{color}
    \definecolor{lightgray}{rgb}{0.95, 0.95, 0.95}
    \definecolor{darkgray}{rgb}{0.5, 0.5, 0.5}
    \definecolor{purple}{rgb}{0.65, 0.12, 0.70}
    \definecolor{blueCode}{rgb}{0, 0, 0.93} 
    \definecolor{greenCode}{rgb}{0, 0.6, 0} 
    \definecolor{redCode}{rgb}{1, 0, 0}